\documentclass[a4paper]{revtex4}
\usepackage{graphicx}
\begin{document}



\title{How fast is the wave function collapse?}
\author{A.~Yu.~Ignatiev}
\email{a.ignatiev@ritp.org}
 \affiliation {{\em Theoretical Physics Research Institute,
     Melbourne 3163,}\\
    {\em   Australia.}}

\begin{abstract}
Using complex quantum Hamilton-Jacobi formulation, a new kind of non-linear equations is 
proposed that have almost classical structure and extend the Schr\"odinger equation to describe the 
collapse of the wave function as a finite-time process. Experimental bounds on the collapse time 
are of order 0.1 ms to 0.1 ps and 
the areas where sensitive probes of the possible collapse dynamics 
can be done include Bose-Einstein condensates, ultracold neutrons or 
ultrafast optics. 
\end{abstract}

\maketitle
 
\section{Introduction}

Recently there has been  considerable interest 
in the fundamental principles of quantum mechanics, including its theoretical,  experimental, and applied aspects \cite{a,l,t},   the physical meaning and reality of the wave function  being a current focus of attention \cite{pbr}. Among the least understood features of quantum mechanics is the collapse of the wave function (also known as the reduction of the wave packet)---``the most interesting point of the entire theory'' \cite{cat}.

In particular, it is not known whether the collapse is instantaneous or takes a finite time \cite{boh}. 

This issue was first raised by von Neumann, and to this day it remains a difficult and unsolved task, notwithstanding the progress in constructing the theories of dynamical collapse \cite{bassi,bb}.  Besides foundational interest, this question is also of utmost importance for quantum computing  and quantum communications \cite{j} because the finite collapse time  could set the ultimate limit to the speed of  quantum computation and information processing. Also, it would shed new light on the basic issues such as the Many-Worlds interpretation of quantum mechanics.
 
Recent years have also seen rapid growth of  attention
to a promising novel approach to quantum mechanics: complex quantum Hamilton-Jacobi (CQHJ) formulation \cite{c}; one of its  main purposes  is   to bring classical and quantum mechanics as close to each other as possible.
An important and interesting question  is whether or not the new formulation can give us new insights into  fundamental long-standing issues such as the quantum measurement problem.

This paper  proposes a new family of `collapsible' Schr\"odinger equations, i.e., non-linear equations describing
 the collapse of the wave function as a dynamical phenomenon with a certain time-scale occurring during the final stages of the measurement process. We then mention briefly  the  experimental consequences of the suggested approach \cite{arx}.
 \section {Classical mechanics in the light of quantum}
Usually, we ask ourselves if quantum mechanics can be 
recast into  a more classical form.
But we can also turn this question upside down and ask if classical mechanics can be given a quantum `look and feel'.

For example, we can ask if anything resembling the wave function $f$ can be introduced into classical mechanics. 

The first obstacle that we meet is this:  If we have two one-particle systems and make out of them one two-particle systems, then wave functions must be {\em multiplied}. But in  classical mechanics, the lagrangians, hamiltonians, or actions must be {\em added}, not multiplied. 

There is an easy way out of this difficulty, though \cite{sr}. We just need to take an additive quantity, like the action $S$, and exponentiate it  thus making the resulting quantity
multiplicative:
\begin{equation}
f=\exp({iS/\hbar}).
\end{equation}
(An arbitrary constant had to be introduced for dimensional reasons, and the natural choice for it is the Planck constant $\hbar$.)

Next, what is the classical `momentum operator'? Recalling that according to the Hamilton-Jacobi theory ${\bf p}=\nabla S$, we obtain:
\begin{equation}
{\bf p}({\bf r},t)=\frac{\hbar}{i}\frac{\nabla f}{f}.
\end{equation}
\section{Derivation of the CQHJ equation} 
We start with the ordinary Schr\"odinger equation for a particle of mass $m$ in the potential $V({\bf r})$:
\begin{equation}
\label{s}
i\hbar\psi_t=-\frac{\hbar^2}{2m}\Delta \psi +V\psi.
\end{equation}
(Throughout the paper, the subscript $t$ denotes the partial derivative with respect to time, $f_t\equiv \partial f/\partial t$.) 
 
Motivated by the previous discussion, let us introduce a new variable ${\bf p}({\bf r}, t)$:
\begin{equation}
\label{p}
{\bf p}=\frac{\hbar}{i}\frac{\nabla \psi}{\psi}.
\end{equation}
Our next goal   is to obtain a closed dynamical equation for ${\bf p}$.
Differentiating Eq. (\ref{p}) with respect to time, we find:
\begin{equation}
\label{4s}
{\bf p}_t=\frac{\hbar}{i}\frac{\nabla \psi_t}{\psi}-{\bf p}\frac{ \psi_t}{\psi}.
\end{equation}
In deriving the second term,  we have used the definition (\ref{p}) rewritten in the form 
\begin{equation}
\label{p1}
\nabla \psi =\frac{i}{\hbar}{\bf p}\psi.
\end{equation}
It is convenient to start $\psi$-elimination from the second term of Eq. (\ref{4s}). For this purpose, we first  find $\Delta \psi$, again using (\ref{p1}):
\begin{equation}
\label{6s}
\Delta  \psi=\nabla\left(\frac{i}{\hbar}{\bf p}\psi \right)=\frac{i}{\hbar}\left(\psi\nabla\cdot{\bf p}+{\bf p}\nabla\psi\right).
\end{equation}
Plugging this into the Schr\"odinger equation (\ref{s}) and multiplying both sides by $i{\bf p}/(\hbar \psi)$, we obtain:
\begin{equation}
\label{11s}
-{\bf p}\frac{ \psi_t}{\psi}=\frac{1}{2m}{\bf p}\nabla\cdot {\bf p}+\frac{i}{2m\hbar}{\bf p} p^2+\frac{i}{\hbar} {\bf p}V.
\end{equation} 
Next, we need to eliminate $\psi$ from the first term of Eq. (\ref{4s}), so we substitute the right-hand side of Eq.  (\ref{6s}) into the Schr\"odinger equation (\ref{s}) instead of $\Delta \psi$, then take the gradient of both parts and divide both of them by $\psi$, which yields:
\begin{equation}
\label{13s}
-i\hbar \frac{\nabla \psi_t}{\psi}=\frac{i\hbar}{2m}\nabla(\nabla\cdot {\bf p})-\frac{1}{2m}{\bf p}\nabla\cdot {\bf p}-\frac{1}{2m}\nabla  p^2-\frac{i}{2\hbar m}p^2 {\bf p}-\nabla V-\frac{i}{\hbar}{\bf p} V.
\end{equation}
We now add the this formula, part-by-part, with Eq. (\ref{11s}) and then use  Eq. (\ref{4s}) to finally obtain:
\begin{equation}
\label{fin}
{\bf p}_t=-\nabla V-\frac{1}{2m}\nabla  p^2+\frac{i\hbar}{2m}\nabla(\nabla\cdot {\bf p}).
\end{equation}
So it turns out that the complete elimination of the wave function is indeed possible, and Eq. (\ref{fin}) gives a new dynamical equation in a closed form.  

This equation plays the role of the Schr\"odinger equation in the CQHJ approach \cite{c}.

A slight rearrangement gives this equation  an elegant form:
\begin{equation}
\label{fin1}
{\bf p}_t=-\nabla(V+\frac{p^2}{2m}+\frac{{\hat{\bf p}}{\bf p}}{2m}),
\end{equation}
where ${\hat{\bf p}}=-i\hbar \nabla$.

This suggests introducing the `quantum Hamiltonian'
\begin{equation}
\label{h}
H=V+\frac{p^2}{2m}+\frac{{\hat{\bf p}}{\bf p}}{2m},
\end{equation}
so we can rewrite our Eq. (\ref{fin}) in a `canonical form'
\begin{equation}
\label{h1}
{\bf p}_t=-\nabla H.
\end{equation}

\section{Non-linear equations for collapse} 
The satisfactory description of the measurement process is likely to require a non-linear extensions of the Schr\"odinger dynamics. If we are looking for such equations, they should combine two features: 
(a) when the quantum system is isolated, they should go over to the standard Schr\"odinger equation;
(b) they should include the non-linear interaction between the system and apparatus.

From the conventional perspective, it is hard to see which form such  equations should take. 
By contrast,  the CQHJ approach can lead us to a solution in the following  way:

First, we replace the condition (a) by an equivalent:
 when the quantum system is isolated, the new equations should go over to the CQHJ equation, which is conveniently rewritten in yet another form 
\begin{equation}
\label{fin1}
{\bf p}_t={\bf F}-\nabla(\frac{p^2}{2m}+\frac{{\hat{\bf p}}{\bf p}}{2m}),
\end{equation}
where  ${\bf F}=-\nabla V$.
Despite being non-linear in terms of $\bf p$, this equation is, of course, linear in terms of $\psi$.

Second, we need to satisfy condition (b). 
A natural way to do that is to add an extra `collapsing' force ${\bf F}_c$ which effectively describes the `measuring' interaction. The key point here is that the extra force must be {\em non-potential}, because otherwise the resulting equation would have the usual Schr\"odinger  form (\ref{s}) with $V$ replaced by $V+V_c$ and therefore would still be linear in $\psi$. 

Thus, finally, one of our non-linear `collapsible Schr\"odinger equations' takes the form 
\begin{equation}
\label{fin2}
{\bf p}_t={\bf F}_c-\nabla H.
\end{equation}
Due to its homogeneity, 
this equation  automatically conserves probability. More precisely, the homogeneity here means the invariance of our equation under the transformation $\psi \rightarrow N(t)\psi $ where $N(t)$ is an arbitrary time-dependent scale factor.

\section{Experimental consequences}
It can be shown \cite{arx} that the collapsible Schr\"odinger equations can lead to remarkable  consequences. Instead of the traditionally assumed instantaneous collapse, we now have the possibility that it is a continuous physical process taking a finite, non-zero time to proceed. This feature offers an exciting opportunity of experimental tests. Due to the limited space,  we refer the interested reader to Ref. \cite{arx} for more details, including the experimental bounds on the collapse time 
(of order 0.1 ms to 0.1 ps) and its convenient dimensionless measure which helps to identify the areas where sensitive probes of the possible collapse dynamics 
can be done. Examples are experiments with Bose-Einstein condensates, ultracold neutrons or 
ultrafast optics. 

Our approach is not to be confused with the well-known  dynamical collapse theories of Ghirardi-Rimini-Weber and Pearle  (CSL) where localization is occurring spontaneously and continuously, even in the absence of a measuring device, due to a stochastic noise the physical source of which is yet to be clarified \cite{bassi}. Unlike CSL, our collapsible  Schr\"odinger equations can be non-stochastic, noiseless. 

Also, it is quite distinct from the earlier models of the Bohm-Bub type \cite{bb} in that it may {\em not} require any hidden variables  \cite{arx}.

The author is thankful to L.  Ignatieva,   V. A. Kuzmin, and M. E. Shaposhnikov for reading the manuscript and helpful discussions. Valuable comments of M. G. Albrow, R. D\"orner, and F. Thaheld are gratefully acknowledged.


\section*{References}
\begin{thebibliography}{<num>}
\bibitem{a}see, e.g.,  S. L. Adler, {\em Quantum Theory as an Emergent Phenomenon} (Cambridge University Press, Cambridge, 2004); G. C. Ghirardi, A. Rimini,  T. Weber,  Phys. Rev. D {\bf 34}, 470 (1986); P. Pearle, Phys. Rev. A {\bf 39}, 2277 (1989); S. Weinberg, arXiv:1109.6462 [quant-ph]; J. S. Bell, Speakable and unspeakable in quantum mechanics (Cambridge University Press, Cambridge, 1989); W. H. Zurek, Physics Today {\bf 44},  36 (1991); E. Joos and H. D. Zeh, Z. Phys.  B {\bf 59}, 223 (1985);
A. E. Allahverdyan, R. Balian, and T. M. Nieuwenhuizen,  arXiv:1107.2138 [quant-ph]; F. H. Thaheld, BioSystems {\bf 92}, 114 (2008);
A. Yu. Ignatiev, Rad. Phys. Chem.  {\bf 75}, 2090 (2006); for a comprehensive bibliography see A. Cabello arXiv: quant-ph/0012089.
\bibitem{l}A. J. Leggett, Science {\bf307}, 871 (2005); J. Phys.: Cond. Mat. {\bf14}, R415 (2002).
\bibitem{t}G. 't Hooft, Class. Quant. Gravity, {\bf 16}, 3263 (1999); AIP Conf. Proc. {\bf 957}, 154 (2007).
\bibitem{pbr} R. Colbeck and R. Renner, Phys. Rev. Lett. 108, 150402 (2012); M. F. Pusey, J. Bartlett, and T. Rudolph,  Nature Physics, doi:10.1038/nphys2309.
\bibitem{cat}E. Schr\"{o}dinger, Naturwissenschaften, 23, 807 (1935), English translation in: J. A. Wheeler and W. H. Zurek  (eds.) {\em Quantum Theory and Measurement}  (Princeton University Press, Princeton, N. J., 1983).

\bibitem{boh}N. Bohr, Nature  {\bf 121}, 580 (1928); J. von Neumann, {\em Mathematische Grundlagen der Quantenmechanik} (Julius Springer-Verlag, Berlin, 1932) [English transl.: Princeton University Press, Princeton, N.J., 1955]; V. B. Braginsky and F. Ya. Khalili, {\em Quantum Measurement} (Cambridge University Press, Cambridge, 1992).
\bibitem{bassi}For a recent review see, e.g., A. Bassi {\em et al.,} arXiv:1204.4325 [quant-ph].
 \bibitem{bb}
D. Bohm and J. Bub, 
Rev. Mod. Phys. {\bf 38}, 453 (1966); 
J. H. Tutsch, Rev. Mod. Phys. {\bf 40}, 232 (1968); 
P. M. Pearle, 
Phys. 
Rev. D {\bf 13}, 857 (1976);  
arXiv:quant-ph/0611211, 0611212 and references therein;
A. N. Grigorenko, J. Phys. A: Math. Gen. {\bf28}, 1459 (1995); T. Durt, Helv. Phys. Acta {\bf72}, 356 (1999).

\bibitem{j}M. A. Nielsen and I. 
L. Chuang, {\em Quantum computation and quantum 
information} (Cambridge University Press, Cambridge, 2010);
R. Jozsa, arXiv:quant-ph/0508124.
\bibitem{c}R. A. Leacock and M. J. Padgett, Phys. Rev. Lett. {\bf 50}, 3 (1983); Phys.
Rev. D {\bf28}, 2491 (1983);
M. V. John, Found. Phys. Lett.,
 {\bf 15}, 329, (2002); Ann. Phys. {\bf324}, 220 (2009); arXiv:1104.3197;
  C.-D. Yang, Ann. Phys. {\bf319}, 399 (2005); {\em ibid}. {\bf319}, 444 (2005);
   Y. Goldfarb, I. Degani, and D. J. Tannor, J. Chem. Phys. {\bf125}, 231103
(2006); 
{\'A}. S. Sanz, F. Borondo, and S. Miret-Art{\'e}s, J. Phys.: Cond. Mat.
{\bf14}, 6109  (2002); 
C. C. Chou {\em et al.,} Phys. Rev. Lett. {\bf102}, 250401 (2009); Ann. Phys. {\bf325}, 2193 (2010); 
B. Poirier,  Phys. Rev. A {\bf77}, 022114 (2008);
C. C. Chou and R. E. Wyatt,  Phys. Lett. A, {\bf374}, 2608 (2010).
\bibitem{arx}A. Yu. Ignatiev, arXiv:1204.3373 [quant-ph].
\bibitem{sr}E. Schr\"odinger, Ann. Physik, {\bf 79}, 361 (1926).
\end{thebibliography}
\end{document}